\documentclass{ws-mpla}

\usepackage[super]{cite}
\usepackage{xcolor}
\usepackage[verbose,hypertexnames=false]{hyperref}
\definecolor{blue}{rgb}{0.0, 0.0, 1.0}
\definecolor{red}{rgb}{1.0, 0.0, 0.0}
\definecolor{royalblue}{rgb}{0.0, 0.14, 0.4}
\hypersetup{colorlinks=true,linkcolor=red, citecolor=blue, urlcolor=blue}
\def\orcid#1{\kern .08em\href{https://orcid.org/#1}{\includegraphics[keepaspectratio,width=0.7em]{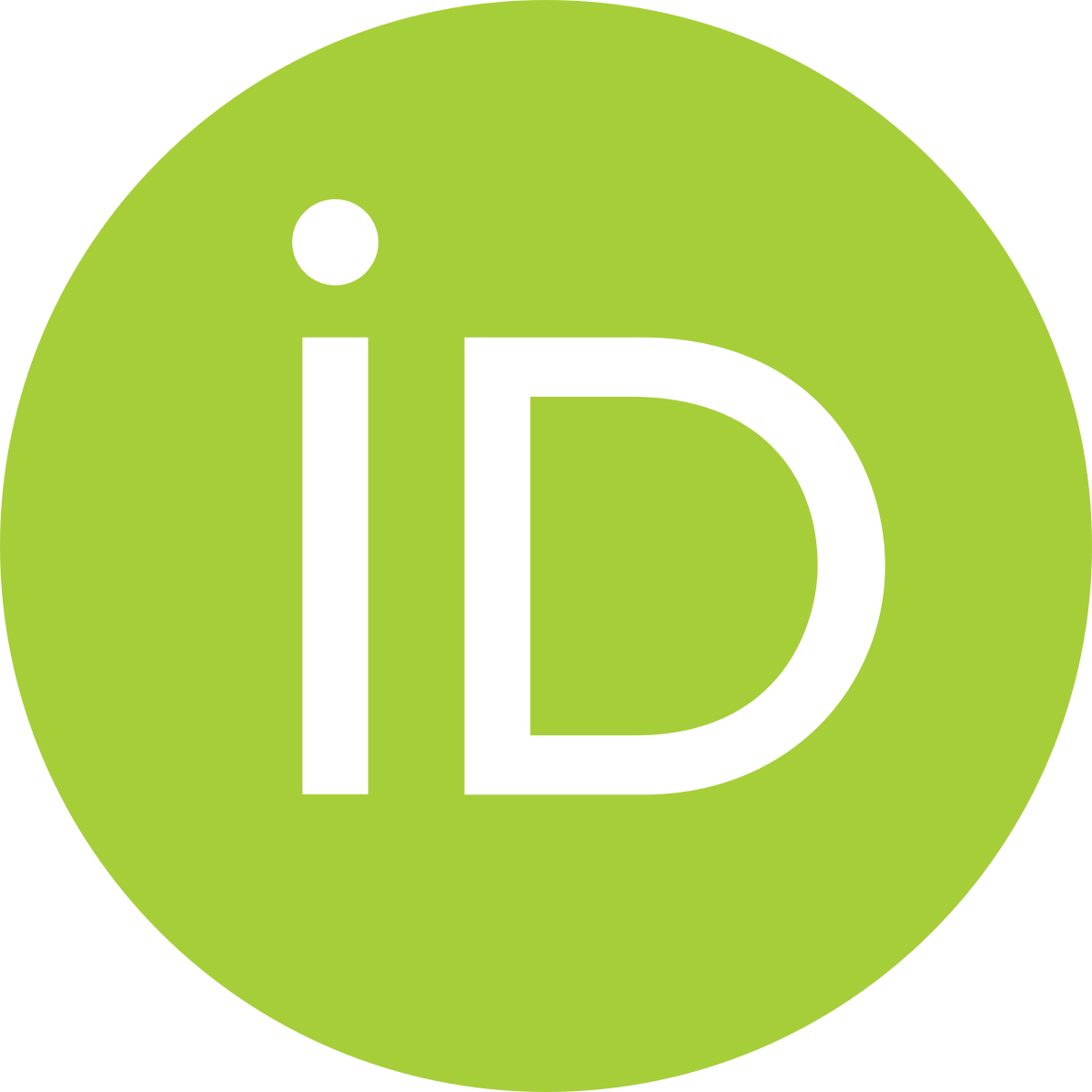}}}

\begin{document}

\markboth{Parada T.~P.~Hutauruk}{Quantum chromodynamics and hadron structure}

\catchline{}{}{}{}{}
\title{A Nambu--Jona-Lasinio model of quantum chromodynamics and hadron structure}

\author{Parada T.~P.~Hutauruk\orcid{0000-0002-4225-7109}}
\address{Department of Physics, Pukyong National University (PKNU)\\
Busan, 48513, Korea \\
Departemen Fisika, FMIPA, Universitas Indonesia\\
Depok, 16424, Indonesia\\
E-Mail Address: phutauruk@gmail.com}
\maketitle

\begin{history}
\received{(Day Month Year)}
\revised{(Day Month Year)}
\accepted{(Day Month Year)}
\published{(Day Month Year)}
\end{history}

\begin{abstract}
In this review paper, I present a study of the structure of hadrons computed in the covariant Nambu-Jona-Lasinio model, which serves as the chiral effective quark theory of QCD. I describe how the NJL model is treated to imitate the spontaneous chiral symmetry breaking and confinement properties of QCD. The consistency for the parton distribution functions and electromagnetic form factors, as internal structure observables, in comparison with existing data and other theoretical predictions, is also shown. The implications of mimicking the QCD properties in the NJL model for hadron structure observables, as well as the relevance of the results to EIC, EicC, and COMPASS/AMBER++ future experiments, are discussed.
\end{abstract}


\section{Introduction}	
Hadron structure is one of the playgrounds of strong interaction of quantum chromodynamics (QCD), in addition to the phase diagram of QCD. It is believed that a greater insight into the hadron structure (in terms of the quarks and gluons degrees of freedom) will lead us to a better understanding of the properties of QCD, like the spontaneously broken chiral symmetry~\cite{Zweig:1964ruk,Zweig:1964jf,Lee:1972fj}, asymptotic freedom~\cite{Politzer:1974fr}, and color confinement or vice versa. Many attempts have been made to study color confinement using various theories or phenomenological approaches~\cite{Shibata:2018tyx,Mandelstam:1974pi,Maldacena:1998zhr,Nambu:1975ba}; however, it remains challenging to completely understand the color confinement of QCD, which remains a long-standing problem at present, although QCD has been around for 50 years~\cite{Gross:2022hyw}. In contrast, the spontaneously broken chiral symmetry (SBCS) has made significant progress, and a better understanding of the SBCS has now emerged. In addition, this SBCS QCD property is obviously captured in the QCD-inspired models, where in the models, it clearly shows that the SBCS mechanism contributes to the dynamical mass of the quark generated by interaction with the vacuum, through the chiral condensate, yielding the emergent hadron mass, excluding the small current quark mass, which is obtained through the Higgs mechanism.

In addition to the theoretical attempts and efforts, several modern experiments, such as the Electron-Ion Collider (EIC)~\cite{Arrington:2021biu}, the Electron-ion collider in China (EicC)~\cite{Anderle:2021wcy}, the COMPASS/AMBER++ at CERN~\cite{Adams:2018pwt}, and the upgraded JLAB 22 GeV~\cite{Accardi:2023chb}, are expected to provide precise enough data and to be promising in providing fundamental physics information on the gluon distribution in hadrons and new insight into QCD properties. This may lead to enhanced understanding of the color confinement of QCD properties, where gluons bind quarks within hadrons. Additionally, these future experiments prompt us to develop a realistic theory that is closely related to quark theory, being close to the QCD mechanism. 

Another attempt from lattice QCD, which is constructed from the first principles of QCD, has been performed. The lattice QCD is a useful tool to delve into information about the quark and gluon dynamics. Significant progress has been made in lattice QCD on hadron structure, including the valence-quark distribution of the pion and nucleon, form factors of the nucleon and pion, the gluon distribution for the pion and nucleon, higher twist contributions, and so forth. It is expected that many more new simulation results will come out from the lattice QCD, which provides complementary physics information to the experimental data. Indeed, the lattice QCD results support us in providing a better understanding of the hadron structure and properties of QCD.

Besides all those attempts, the global QCD analysis tools are also an alternative tool in providing physics information directly obtained from experimental data. In global QCD analysis, the experimental data are fitted to a specific functional form of observables using various statistical techniques to handle the uncertainties. To do so, it needs a large amount of data, besides a satisfactory method. This is due to the fact that as more data is collected, a more accurate model prediction is obtained; therefore, in analysis, the data used are usually collected from various experiments (world data). Recently, in the global analysis, the approach has been extended to use machine learning (ML)~\cite{Allaire:2023fgp}. With the ML technique, it is expected that ML will accelerate the running process and handle the uncertainties more precisely and accurately. However, in this paper, I emphasize that I concentrate more on the theoretical model approaches, namely the NJL model.    

This paper is organized as follows. In Sec.~\ref{sec:qcd}, I briefly introduce the properties of QCD, in particular in the nonperturbative QCD regime, where the scale is closely related to the hadron structure. Additionally, I present how the NJL model captures the SBCS and confinement QCD properties. Section~\ref{sec:struc} presents the expressions of parton distribution functions (PDFs) and electromagnetic form factors (EMFFs). Next, I present our numerical results for the $\pi^+$ and $K^+$ PDFs and EMFFs in Sec.~\ref{sec:num}. Summary and conclusion of this work are devoted to Sec.~\ref{sec:sum}.

\section{Quantum Chromodynamics} \label{sec:qcd}
\begin{figure}[ht]
\centering
\includegraphics[width=0.85\linewidth]{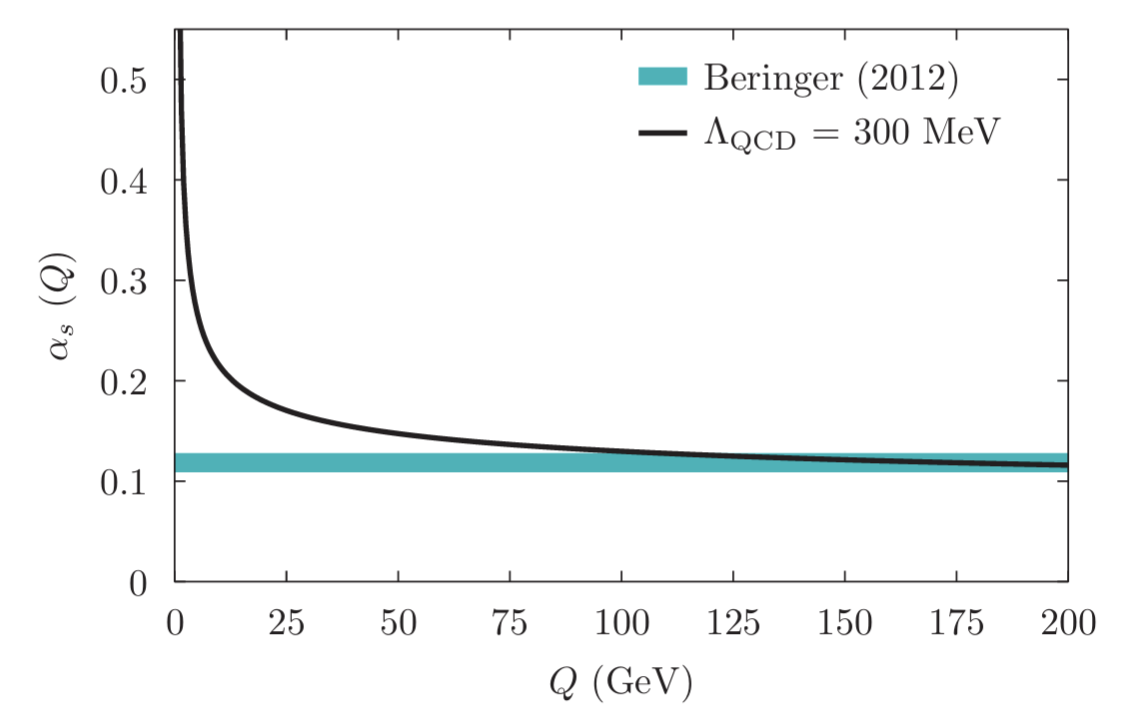}
\caption{Running coupling $\alpha_S (Q)$ as a function of $Q$. The green band represents the prediction value of $\alpha_S (M_z^2 = 0.1184 \pm 0.0007)$, taken from Ref.~\cite{ParticleDataGroup:2012pjm}.}
\label{fig1}
\end{figure}

It is widely known that QCD~\cite{Fritzsch:1973pi}, which is a part of the Standard Model (SM), plays an important role in describing the strong interaction between quarks and gluons forming hadrons. QCD can be split into perturbative QCD and nonperturbative QCD, based on the strong running coupling $\alpha_S (Q^2)$ and scale. Here, I limit the discussion to nonperturbative QCD at low energy, which is the main focus of this work. The nonperturbative QCD is very difficult to treat due to its strong running coupling, which cannot be approached perturbatively. The key features of nonperturbative QCD are color confinement and SBCS. The color confinement is an interesting feature of nonperturbative QCD at large distances. In this regime, the interaction of quarks and gluons is very strong, as specified by its large running coupling, which gradually grows in the infrared regime (at very low $Q$ as clearly indicated in Fig.~\ref{fig1}). This is known as \textit{infrared slavery}. Also, in this regime, the perturbative QCD fails. It is worth noting that nonperturbative QCD has infrared singularities, preventing the quarks and gluons from being released from hadrons.

In the past, several theorists attempted to model the non-Abelian theories by adding the gluon interaction terms into the QCD Lagrangian. However, the developments bring us in an inappropriate direction, which is far away from reality. Later, Gross, Wilczek, and Politzer~\cite{Fritzsch:1973pi,Politzer:1973fx,Gross:1973id} discovered that the effective coupling constant $g_s$ disappears at short distances or large $Q^2$ as depicted in Fig.~\ref{fig1}; since then, that perspective on that direction has completely changed. Such a developed theory paved the way to confinement and bridged between the confinement at low energy and the quasi-free behavior of quarks in the parton model. Oppositely, in the short distances or high energy scale, the QCD has a feature of asymptotic freedom, where the $g_s$ is relatively small. At this distance, the quarks and gluons' interaction decreases, and QCD is expected to be asymptotically free. A formula for $\alpha_S (Q^2)$ is defined as
\begin{eqnarray}
\label{eq1}
    \alpha_S (Q^2) &=& \frac{g_s}{4\pi} = \frac{4\pi}{\beta_0 \ln \big[ \frac{Q^2}{\Lambda_{\mathrm{QCD}}^2}\big]},
\end{eqnarray}
where $\beta_0 = (11 N_c -2N_f)/3$ is a constant with $N_f$ is the number of quark flavors and $N_c=3$ denotes the number of colors and $\Lambda_{\mathrm{QCD}} =$ 200-300 MeV is the QCD scale, depending on the typical hadron size. This value of $\Lambda_{\mathrm{QCD}}$  is consistent with the experimental result.

\begin{figure}[ht]
\centering
\includegraphics[width=0.85\linewidth]{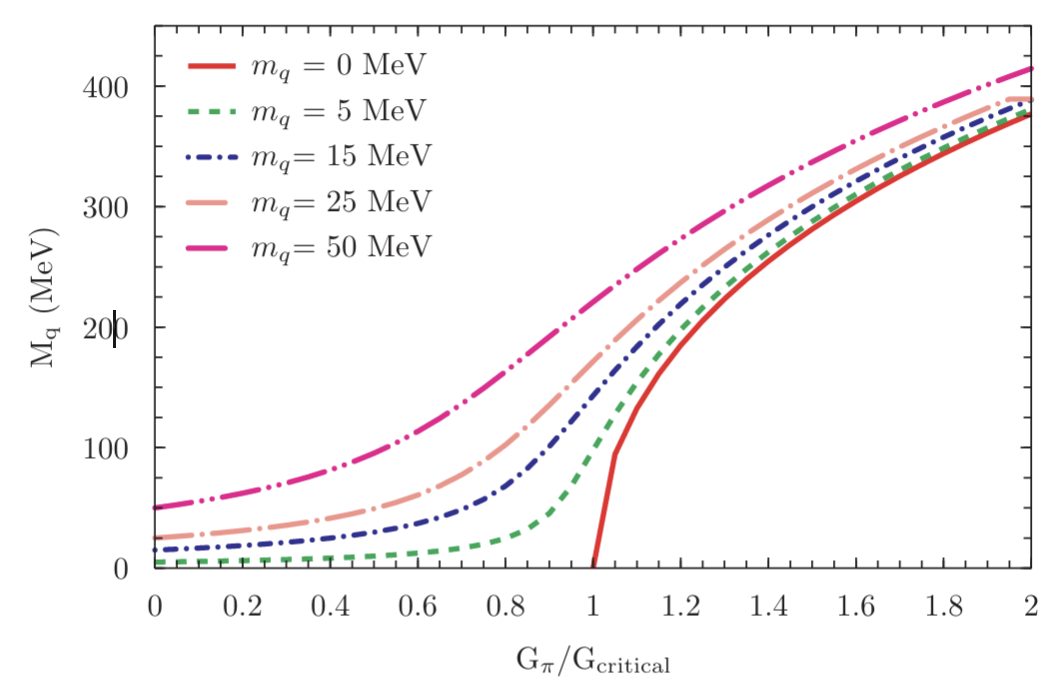}
\caption{Dynamical quark mass generation for different values of the current quark masses as a function of $G_\pi /G_{\mathrm{critical}}$.}
\label{fig2}
\end{figure}

I now turn on another feature of nonperturbative QCD. DCSB is responsible for generating the quark mass of the hadron at a low energy scale, which is commonly below 1 GeV. By taking the quarks to be massless, the chiral symmetry is realized in the Goldstone mode, spontaneously broken. The quarks become asymptotically free when $\alpha_S (Q^2)$ at $Q^2 \rightarrow \infty$. This DCSB feature is clearly captured in the covariant NJL model, as the chiral quark effective theory of QCD, as shown in Fig.~\ref{fig2}, while the confinement is applied through the infrared (IR) cutoff of the Schwinger proper time regularization scheme. Shortly, the expression of the dynamical quark mass in the Schwinger proper time regularization scheme can be given by~\cite{Hutauruk:2016sug}
\begin{eqnarray}
    \label{eq2}
    M_q &=& m_q + \frac{3 G_\pi M_q}{\pi^2} \int_{\tau_{\mathrm{UV}}}^{\tau_{\mathrm{IR}}} \frac{d\tau}{\tau^2} \exp \big[ -\tau (M_q^2)\big],
\end{eqnarray}
where $\tau_{\mathrm{UV}} = 1/\Lambda_{\mathrm{UV}}^2$ and $\tau_{\mathrm{IR}} = 1/\Lambda_{\mathrm{IR}}^2$ are the ultraviolet (UV) and IR limits with $\Lambda_{\mathrm{UV}} =$ 645 MeV and $\Lambda_{\mathrm{IR}} =$ 240 MeV are respectively UV and IR cutoffs. The values of $M_q =$ 400 MeV are set. Details of the parameter determination can be found in Ref.~\cite{Hutauruk:2016sug,Hutauruk:2018zfk}.

Figure~\ref{fig2} shows that the trivial solution of $M_q$ is given when $G_\pi =0$, while the nonperturbative solution emerges at a nonzero value of $G_\pi$. It is worth noting that when $m_q =$ 0 gives $m_\pi =$ 0, which corresponds to the Goldstone theorem, and the Nambu-Goldstone phase happens when the chiral symmetry is dynamically broken. In the chiral limit ($m_q \sim 0$), $M_q$ has nontrivial solution, where $M_q \neq 0$ yielded $G_\pi > G_{\mathrm{critical}}$. Such a phenomenon is related to the DCSB and the Nambu-Goldstone phase. Also, in the chiral limit, the critical coupling is given by $G_{\mathrm{critical}} = \pi^2/[3 (\Lambda^2_{\mathrm{UV}} - \Lambda^2_{\mathrm{IR}})]$. This clearly shows that the dynamical mass generation is associated with the chiral condensate generation. Moreover, in the chiral limit, it shows that the chiral condensate is zero at $G_\pi < G_{\mathrm{critical}}$, which is known as the Wigner-Weyl phase. Here, I have demonstrated that the NJL model shows a clear DCSB mechanism, \textcolor{blue}{which is one of the features of QCD}.

A color confinement, another feature of nonperturbative QCD, is worth mentioning that the NJL model is non-renormalizable and not a confining model. Therefore, in the NJL model, some regularization schemes to cure the divergences must be applied. In the literature, there are some possible regularization schemes for tackling this problem. Here, among others, I choose the Schwinger proper time regularization scheme by considering the IR and UV cutoffs in the calculation. These cutoffs are important to obtain finite solutions. The generic expression of the Schwinger proper time regularization scheme is given by
\begin{eqnarray}
    \label{eq4}
    \frac{1}{X^n} &=& \frac{1}{\big(n-1\big)!} \int_{\tau_{\mathrm{UV}}}^{\tau_{\mathrm{IR}}} d\tau \tau^{[n-1 ]} \exp \big[ -\tau \big( X \big)],
\end{eqnarray}
where $X^n$ is the denominator obtained after applying the Feynman parameterization and Wick rotation to rotate the loop integral to Euclidean space. Note that in practice, the UV cutoff is required to get the theory finite, but in a quark bound state, the IR cutoff is also introduced. The IR cutoff is useful to eliminate the nonphysical quark threshold for the hadrons to decay into free quarks. This IR cutoff has to be introduced to reflect the QCD confinement scale. 

After explaining how the DCSB and color confinement, which are features of nonperturbative QCD, are treated and captured in the NJL model. Next, I present how to demonstrate the NJL model to compute the PDFs and EMFFs of hadrons in Sec.~\ref{sec:struc}.

\section{Hadron Structure} \label{sec:struc}
Here, I present the approaches employed and observables observed to study the internal structure of hadrons. The hadron structure is encapsulated in the parton distribution function (PDFs), generalized parton distributions (GPDs), electromagnetic form factors (EMFFs), fragmentation functions (FFs), gravitational form factors (GFFs), and parton distribution amplitudes (PDAs). In this work, I will concentrate on the computations of the PDFs and EMFFs of the $\pi^+$ and $K^+$ in the chiral effective quark theory, for example, the covariant NJL model. Hereafter, I will specifically discuss the $\pi^+$ and $K^+$ PDFs and EMFFs in the NJL approach. It is worth mentioning that in the literature, several models, which are constructed in the same manner, have also been used to investigate the PDFs and EMFFs or their hadron structure observables~\cite{Maris:2000sk,Hernandez-Pinto:2023yin,Abidin:2019xwu,Hutauruk:2025wkn,Lan:2019rba,Liu:2023yuj}. Among them, for example, is the Dyson Schwinger Equation (DSE) model\cite{Maris:2000sk}. This model has a different treatment from the NJL model. This DSE model has quark momentum-dependent terms in the quark propagator, whereas the NJL model does not depend on the quark momentum (see further about the DSE model in Ref.~\cite{Roberts:1994dr}).

In the NJL model~\cite{Klevansky:1992qe}, the expressions of the PDFs of the $\pi^+$ and $K^+$ in the Schwinger proper time regularization scheme can be shortly defined as~\cite{Hutauruk:2016sug}
\begin{eqnarray}
    \label{eq5}
    q_{PM} (x) &=& \frac{N_c g_{M qq}^2}{4\pi^2} \int \frac{d\tau}{\tau} \exp \big[-\tau \big(  k^2 (x^2 -x) + xM_2^2 -M_1^2 (x-1) \big) \big], \nonumber \\
    &\times& \big[ 1+\tau \big(k^2 (x-x^2) -(x-x^2) (M_2-M_1)^2 \big)\big],
\end{eqnarray}
\begin{eqnarray}
    \label{eq5}
    \bar{q}_{PM} (x) &=& \frac{N_c g_{M qq}^2}{4\pi^2} \int \frac{d\tau}{\tau} \exp \big[-\tau \big(  k^2 (x^2 -x) + xM_1^2 -M_2^2 (x-1) \big) \big], \nonumber \\
    &\times& \big[ 1+\tau \big(k^2 (x-x^2) -(x-x^2) (M_1-M_2)^2 \big)\big],
\end{eqnarray}
where the subscript of $PM =(\pi^+, K^+)$ stands for the pseudoscalar mesons, $M_1$, $M_2$ are respectively the quark and antiquark dynamical quark masses and $N_c =3$ is the quark number of colors. Note that for the $\pi^+$ case, one takes $M_1 =M_2 =M$ and replaces the $g_{K qq} \rightarrow$ $g_{\pi qq}$. Also, the valence quark distribution can be defined as $q_v (x) = q (x) - \bar{q} (x)$. Note that the $\pi^+$ and $K^+$ PDFs should satisfy the baryon number and momentum sum rules (conservation rules) are defined as $\int_0^1 dx u_v^K (x) =1 = \int_0^1 dx \bar{s}_v^K (x)$ and $\int_0^1 dx x \big[ u_v^K + \bar{s}_v^K (x) \big] = 1$, respectively.

Besides the $\pi^+$ and $K^+$ PDFs, I also show the expressions of the $\pi^+$ and $K^+$ EMFFs calculated in the NJL model. The  formula for the $\pi^+$ and $K^+$ dressed EMFFs in the Schwinger proper time regularization scheme is given as follows~\cite{Hutauruk:2016sug}
\begin{eqnarray}
\label{eq6}
    F_{PM} (Q^2) &=& \big[ F_{1U} (Q^2) f_{PM}^{us} (Q^2) -F_{1S} (Q^2) f_{PM}^{su} (Q^2) \big] \nonumber \\
    &+& \big[ F_{2U} (Q^2) - F_{2S} (Q^2) \big] f^T_{PM} (Q^2), 
\end{eqnarray}
where the $f_{PM}^{us} (Q^2)$ and $f_{PM}^{su} (Q^2)$ are respectively defined as 
\begin{eqnarray}
    \label{eq7}
    f_{PM}^{us} (Q^2) &=& \frac{N_cg_{PM qq}^2}{4\pi^2} \int_0^1 dx \int \frac{d\tau}{\tau}  \exp \big[ -\tau \big( q^2 (x^2-x) + M_1^2\big)\big] \nonumber \\
    &+& \frac{N_c g_{PM qq}^2}{4\pi^2} \int_0^1 dx \int d\tau \nonumber \\
    &\times& \exp \big[ k^2 ((x+z)^2 -(x+z)) -xz q^2 + M_2^2 (1-x-z) + M_1^2 (x+z) \big] \nonumber \\
    &\times& \big[ k^2 (x+z) + (M_1-M_2)^2 (x+z) -2M_2^2 + 2 M_2 M_1\big],
\end{eqnarray}
\begin{eqnarray}
    \label{eq7}
    f_{PM}^{su} (Q^2) &=& \frac{N_cg_{PM qq}^2}{4\pi^2} \int_0^1 dx \int \frac{d\tau}{\tau}  \exp \big[ -\tau \big( q^2 (x^2-x) + M_2^2\big)\big] \nonumber \\
    &+& \frac{N_c g_{PM qq}^2}{4\pi^2} \int_0^1 dx \int d\tau \nonumber \\
    &\times& \exp \big[ k^2 ((x+z)^2 -(x+z)) -xz q^2 + M_1^2 (1-x-z) + M_2^2 (x+z) \big] \nonumber \\
    &\times& \big[ k^2 (x+z) + (M_2-M_1)^2 (x+z) -2M_1^2 + 2 M_1 M_2\big],
\end{eqnarray}
and the quark sector form factors are given as
\begin{eqnarray}
    \label{eq8}
    F_{1U}(Q^2) &=& \frac{1}{2} \big[ F_{1\omega} (Q^2) + F_{1\rho} (Q^2)\big], \\
    F_{1D}(Q^2) &=& \frac{1}{2} \big[ F_{1\omega} (Q^2) - F_{1\rho} (Q^2)\big], \\
    F_{1S}(Q^2) &=&  F_{1\phi} (Q^2).
\end{eqnarray}
The expression for the dressed quark form factors can be defined by
\begin{eqnarray}
    \label{eq9}
    F_{1i} (Q^2) &=& \frac{1}{1 +2 G_i \Pi_{vv} (Q^2)}, \hspace{2cm} F_{2i} (Q^2) =0.
\end{eqnarray}
Details of the derivations and final expressions for the $\pi^+$ and $K^+$ PDFs and EMFFs can be found in Ref.~\cite{Hutauruk:2016sug}. Analogous to the PDFs of $\pi^+$ and $K^+$ cases, the $K^+$ EMFFs in Eq.(\ref{eq6}) are straightforwardly applied to that for the $\pi^+$ case by replacing the $g_{K qq} \rightarrow$ $g_{\pi qq}$ and taking $M_1 = M_2 =M$. Using these PDFs and EMFFs, I will present the numerical results in Sec.~\ref{sec:num}.

\section{Numerical Result and Discussion} \label{sec:num}
\begin{figure}[ht]
\centering
\includegraphics[width=0.485\linewidth]{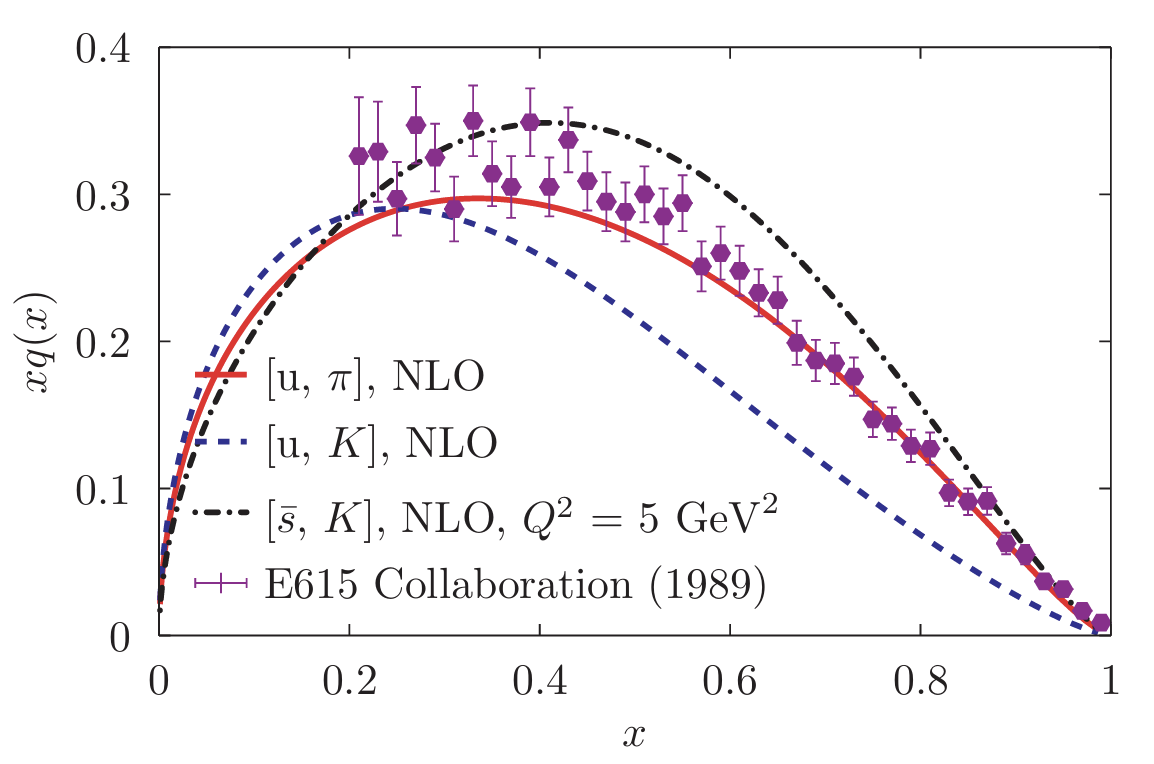}
\includegraphics[width=0.485\linewidth]{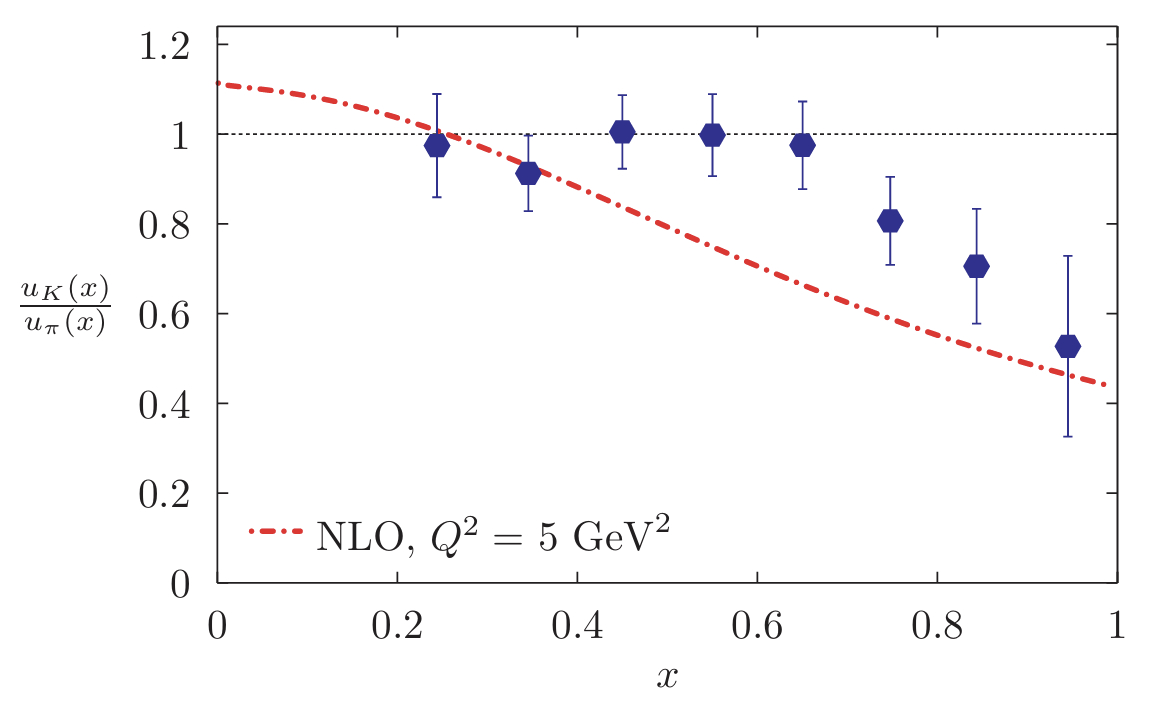} 
\caption{Results for the $\pi^+$ and $K^+$ PDFs at $Q^2 =$ 5 GeV$^2$, evolved from the initial scale $Q_0^2 =$ 0.16 GeV$^2$ using the NLO DGLAP QCD evolution~\cite{Miyama:1995bd}. The $xu_\pi (x)$, $xu_K (x)$, and $xs_K (x)$ as a function of the quark longitudinal momentum $x$ (left panel), and the ratio of the $u_K (x)/u_\pi (x)$ as a function of $x$ (right panel). Experimental data is taken from Ref.~\cite{E615:1989bda}.}
\label{fig3}
\end{figure}

\begin{figure}[ht]
\centering
\includegraphics[width=0.485\linewidth]{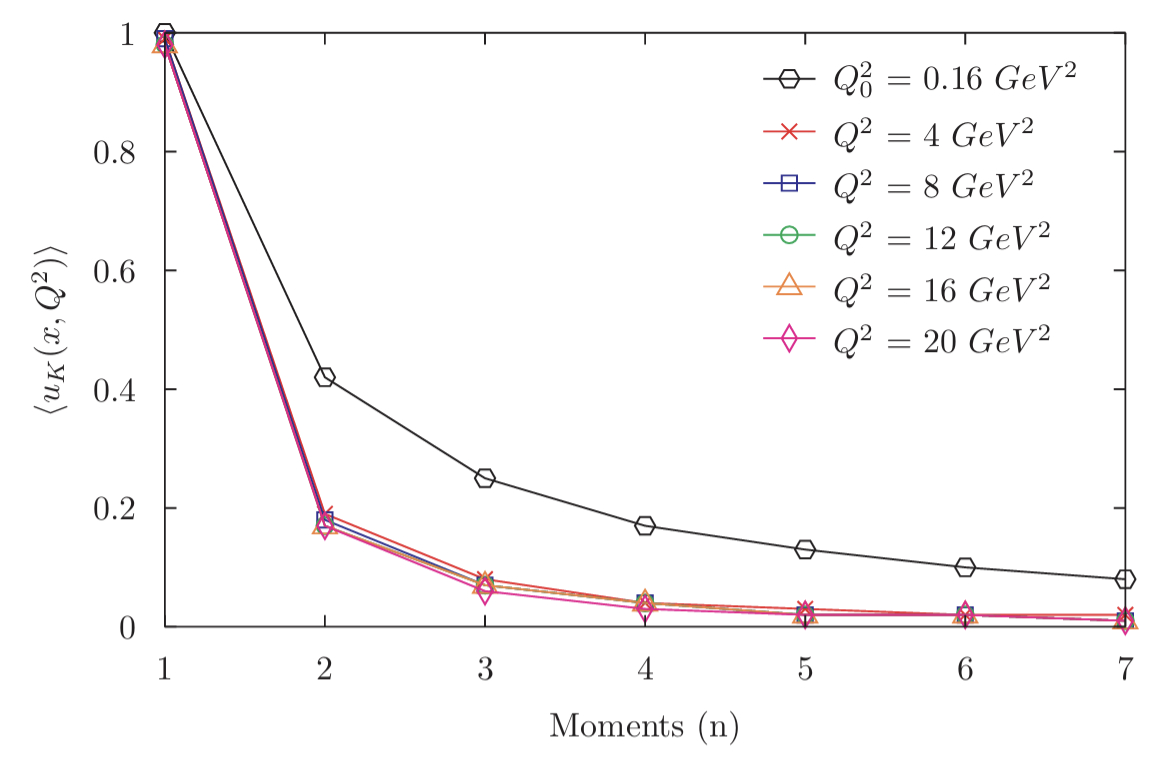}
\includegraphics[width=0.485\linewidth]{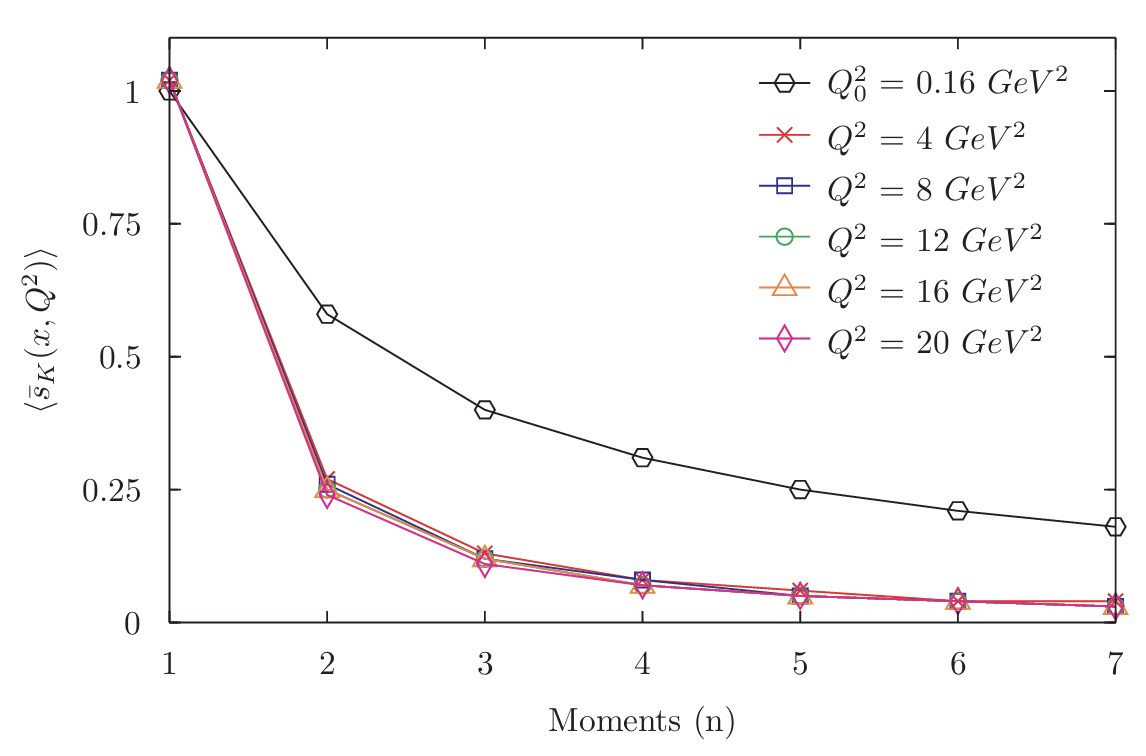} \\
\includegraphics[width=0.485\linewidth]{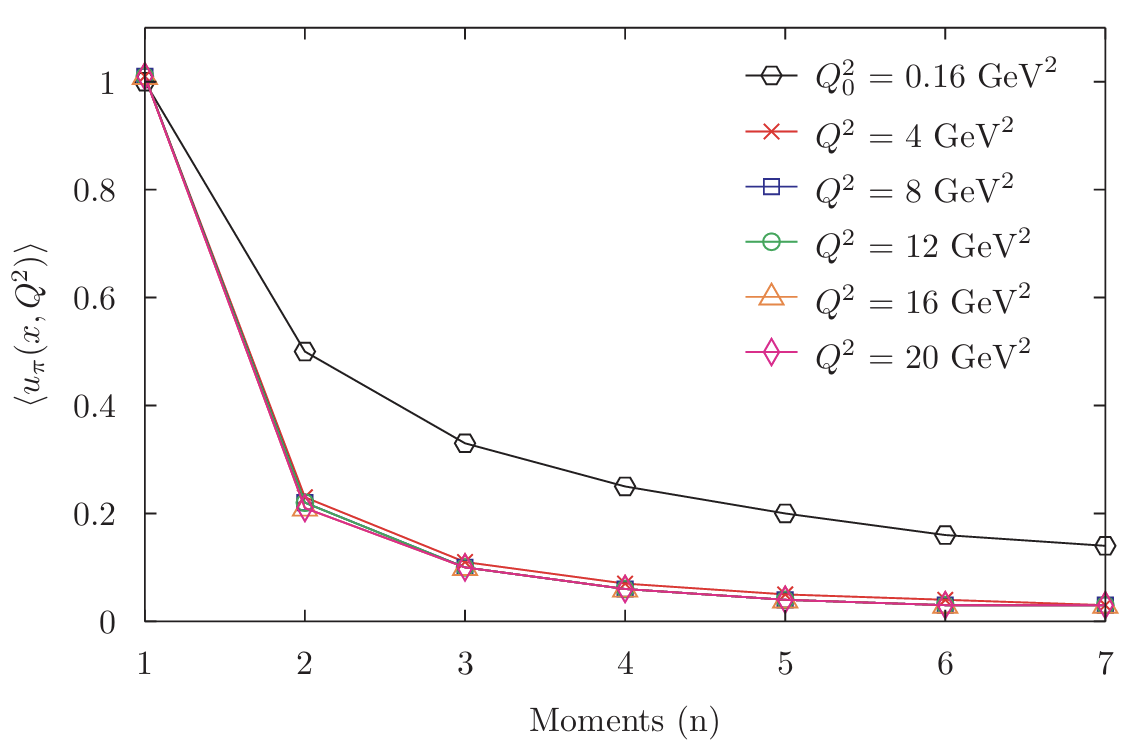}
\caption{The moments of $K^+$ and $\pi^+$ PDFs for different renormalization scales $Q^2 =$ 0.16, 4, 8, 12, 16, and 20 GeV$^2$. Note the moments are calculated using $\big< x^{n-1} \big> = \int_0^1 dx x^{n-1} q_v (x) $, where $n=1, 2, \cdot \cdot \cdot$ is integer, and for $n=1$ (first moment) is the PDFs normalization, $\int_0^1 dx q_v (x) = 1$.}
\label{fig4}
\end{figure}

\begin{figure}[ht]
\centering
\includegraphics[width=0.485\linewidth]{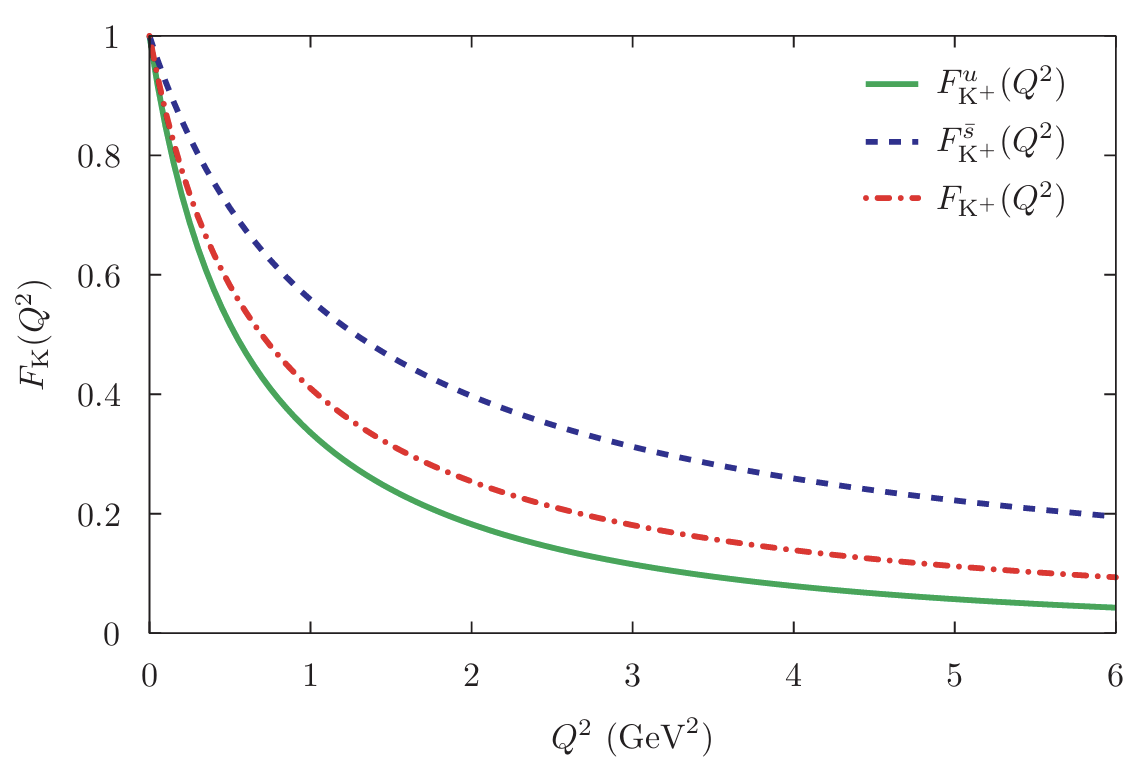}
\includegraphics[width=0.485\linewidth]{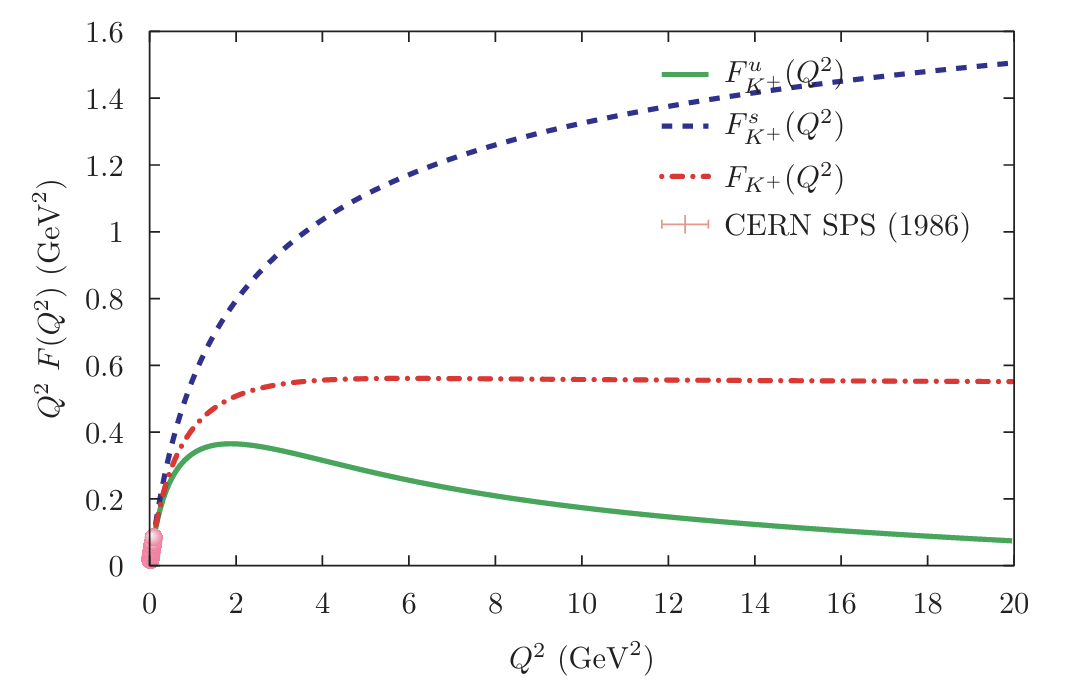} \\
\includegraphics[width=0.485\linewidth]{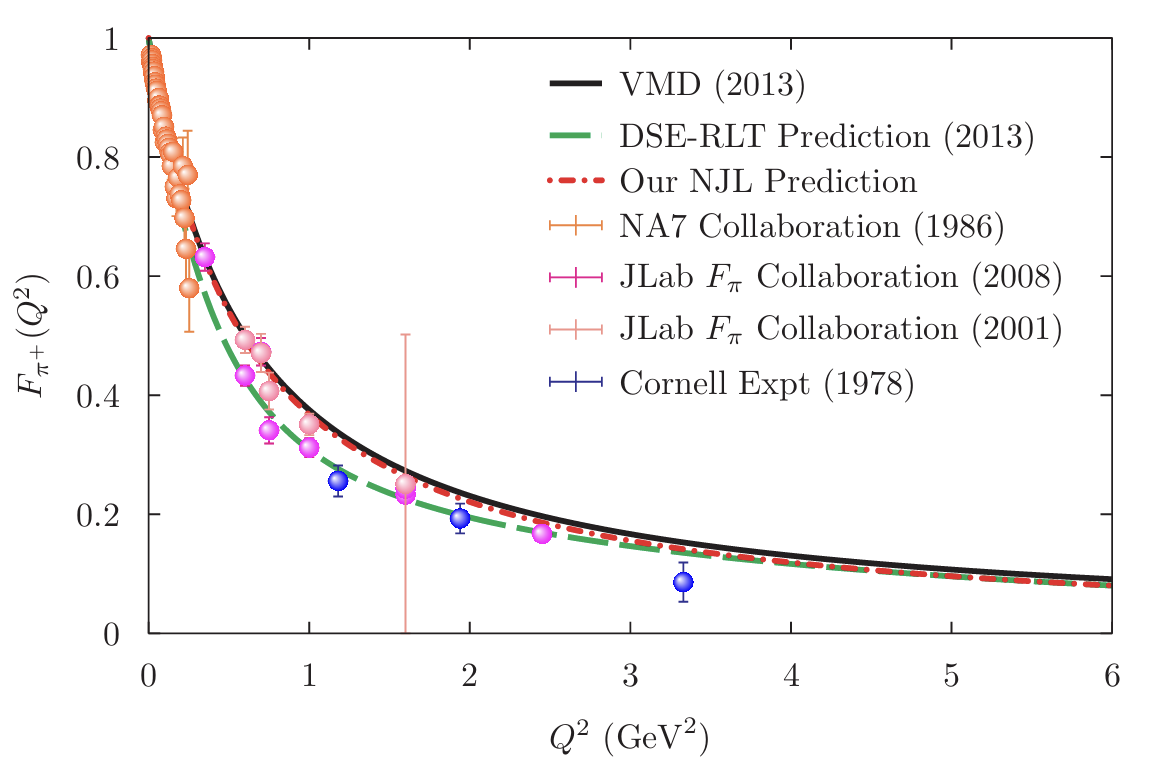}
\includegraphics[width=0.485\linewidth]{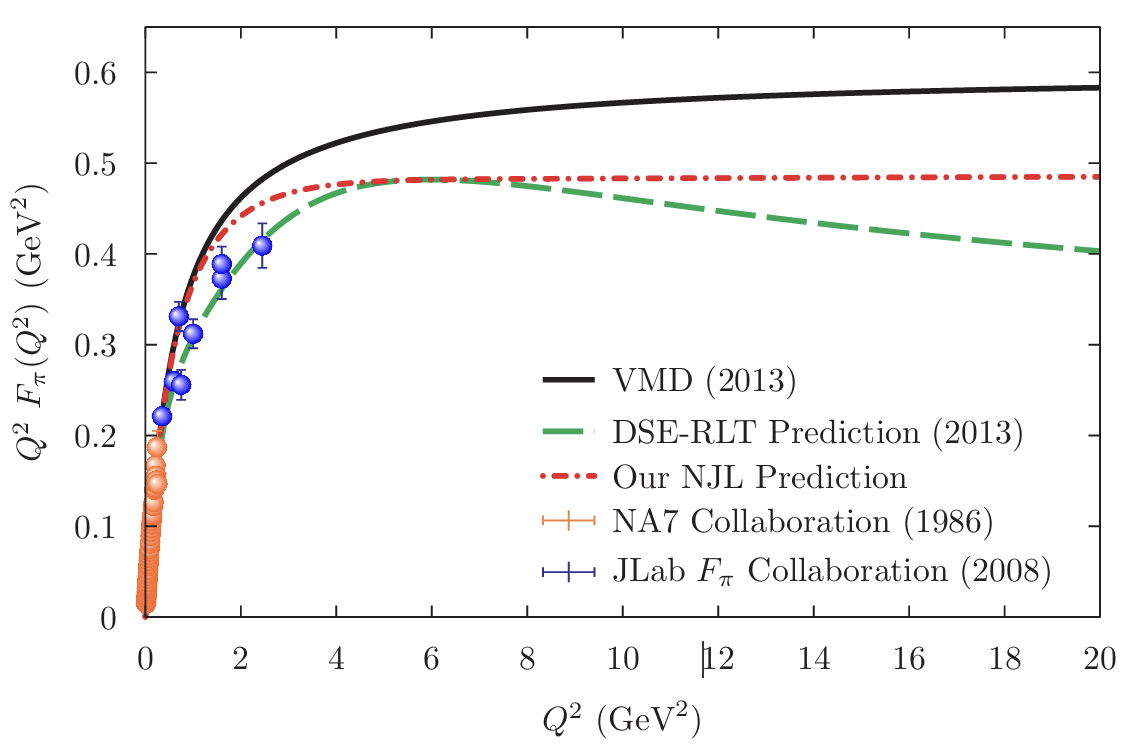}
\caption{Results for the dressed EMFFs of the $K^+$ and its quark sector form factors (upper left panel), the EMFFs of the $K^+$ and their quark sector form factor multiplying with $Q^2$ in comparison with data (low $Q^2)$~\cite{Amendolia:1986ui} (upper right panel), the EMFFs of the $\pi^+$ in comparison with the data~\cite{Amendolia:1984nz,NA7:1986vav,JeffersonLabFpi-2:2006ysh,JeffersonLabFpi:2007vir,JeffersonLab:2008jve,JeffersonLab:2008gyl} and other theoretical models (lower left panel), and the EMFFs of the $\pi^+$ and others multiplying with $Q^2$ (lower right panel).}
\label{fig5}
\end{figure}

Here, the numerical results for the $\pi^+$ and $K^+$ PDFs and EMFFs are presented in Figs.~\ref{fig3}-\ref{fig5}, and their implications are discussed.

The left panel of Fig.~\ref{fig3} shows the valence quarks of the $\pi^+$ and $K^+$ at $Q^2 =$ 5 GeV$^2$ in comparison to the E615 data~\cite{E615:1989bda}. It is found that the up valence-quark distributions of the $\pi^+$ are nicely fit with the data~\cite{E615:1989bda}. Unfortunately, the data for the valence-quark distributions of the $K^+$ are sparse at present; therefore, the comparison of the $K^+$ valence quark distributions with the data cannot be shown. However, several attempts have been planned to measure the $\pi^+$ and $K^+$ PDFs at the EIC~\cite{Arrington:2021biu}, the EicC~\cite{Anderle:2021wcy}, the COMPASS/AMBER++~\cite{Adams:2018pwt}, and the upgraded JLAB 22 GeV~\cite{Accardi:2023chb}. At that time, it becomes possible to confront the PDF results of this work and other model predictions with the new precise data.

In the right panel of Fig.~\ref{fig3}, I present the ratio of $u_K (x)/u_\pi (x)$ at $Q^2 =$ 5 GeV$^2$. It shows that the prediction results of the $u_K(x)/u_\pi (x)$ are in good agreement with the data, in particular at $ 0.2<x<0.4$ and around $x \simeq 0.95$. It is worth noting that, in the literature, some theoretical models, such as the Dyson-Schwinger equation (DSE), show good predictions for this ratio data. However, there is no rigorous conclusion made to this ratio result; therefore, again, new precise data is absolutely required to make a rigorous conclusion and to select the best theoretical model to describe the data.  

In addition, in the upper left of Fig.~\ref{fig4}, I show the results for the moments of $\big< x^{n-1} \big>$ for the $u_K (x)$ with different renormalization scales $Q^2$. It is found that the moments of the $u_v^K(x)$ decrease with increasing $n$, but it seems the values of $\big<x^{n-1} \big>$ do not significantly change with increasing values of the renormalization scale $Q^2$. In other words, the values of the moment are insensitive to the renormalization scale $Q^2$. These behaviors are followed by the moments for the $\bar{s}_K (x)$ and $u_\pi (x) $ in the right upper and lower center panels of Fig.~\ref{fig4}. However, the magnitudes of moments for the $\bar{s}_K (x)$  at scale $Q^2=Q_0^2 =$ 0.16 GeV$^2$ and for all values of $Q^2$ are bigger compared to those for the $u_K (x)$ for the corresponding renormalization scales. It can be understood since the strange quark carries more momentum of the $K^+$ parent. Such calculations have been done in various theoretical models available in the literature. It is worth mentioning that lattice QCD simulations are only able to compute a few moments of the $\pi^+$ and $K^+$ PDFs. It is expected that the higher moments computation results of the PDFs for the $\pi^+$ and $K^+$ from the lattice QCD will be available in the future. Results for the gluon distributions, which are relevant to the EIC physics program, for the $\pi^+$ and $K^+$ from the lattice and theoretical models, have been available in the literature. These attempts are, of course, essential for a better understanding of the hadron structure and the features of QCD. 

Next, in Fig.~\ref{fig5}, the EMFFs of the $\pi^+$ and $K^+$ with their quark sector form factors, as well as their multiplication with $Q^2$, are shown. In the upper left panel of Fig.~\ref{fig5}, it shows the $K^+$ EMFFs and their quark sector form factors decrease as the $Q^2$ increases, as expected. Such results are also found by other theoretical and lattice QCD simulation results. To clearly show the features of the $K^+$ EMFFs, I show the results for $Q^2 F_K (Q^2)$ and its quark sector form factors in the upper right panel of Fig.~\ref{fig5}. I find that the $Q^2 F_K (Q^2)$ at higher $Q^2$ (asymptotic/conformal region) satisfies the quark counting rules $\simeq 1/Q^2$, while $Q^2 F^s_K (Q^2)$ increases and $Q^2 F^u_K (Q^2)$ decreases with change of $Q^2$. Also, it is shown that the $K^+$ EMFFs are consistent with low $Q^2$ data.

The lower left panel of Fig.~\ref{fig5} shows the results for the $\pi^+$ EMFFs in comparison with the data~\cite{Amendolia:1984nz,NA7:1986vav,JeffersonLabFpi-2:2006ysh,JeffersonLabFpi:2007vir,JeffersonLab:2008jve,JeffersonLab:2008gyl} and other theoretical model results. It is found that the $\pi^+$ EMFFs are consistent with the data and other model results. In the lower right panel of Fig.~\ref{fig5}, it shows that results for $Q^2 F_\pi (Q^2)$ at higher $Q^2$, having a power counting rule $\simeq 1/Q^2$.
However, it is rather different from the DSE prediction results at around $Q^2 \simeq$ 6 GeV$^2$. This is due to the momentum dependence is absent in our approach. Note that the availability of the $K^+$ EMFFs data is limited, only covering a very small $Q^2$; therefore, more precise data for the $\pi^+$ and $K^+$ EMFFs are urgently needed at higher $Q^2$, so the behavior of the $\pi^+$ and $K^+$ EMFFs can be concluded not only at low $Q^2$ but also at higher $Q^2$ (asymptotic regime).

\section{Summary and conclusion} \label{sec:sum}
As a summary, in this paper, I have explained the construction of the NJL model with the Schwinger proper time regularization scheme, which successfully mimics the QCD properties, i.e, confinement (even though in a crude way) and DCSB. I then show the NJL model prediction on the $\pi^+$ and $K^+$ PDFs and EMFFs. 

Overall, results for the $\pi^+$ and $K^+$ PDFs and EMFFs are consistent with the data and other models. The results of this work are promising to confront the new precise data from future experiments, such as the EIC~\cite{Arrington:2021biu}, the EicC~\cite{Anderle:2021wcy}, the COMPASS/AMBER++~\cite{Adams:2018pwt}, and the upgraded JLAB 22 GeV~\cite{Accardi:2023chb}.

From a theoretical side, the theory developments with various approaches for the calculations of the hadron PDFs and EMFFs, as well as other observables, for different reaction processes, such as the Sullivan process, the Drell-Yan process, and other exclusive hard processes, deserve further studies and investigations. Additionally, the gluon distribution of the hadrons is also a crucial quantity to study, since it is one of the main physics programs of EIC and is closely connected to the color confinement.  

Finally, it is also worth mentioning that the results from the lattice QCD and global analysis for the $\pi^+$ and $K^+$ PDFs and EMFFs (other hadrons and observables), and the improved approaches and strategies, are crucially needed 
for a better understanding of the internal structure of the $\pi^+$ and $K^+$ Goldstone boson.

\section*{Acknowledgments}
This work was partially supported by the PUTI Q1 Research Grant from the University of Indonesia (UI) under contract No. NKB 442/UN2.RST/HKP.05.00/2024 and the RCNP Collaboration Research Network program under project number COREnet 057.

\section*{ORCID}
\noindent Parada T.~P.~Hutauruk \orcid{0000-0002-4225-7109} \url{https://orcid.org/0000-0002-4225-7109}


\end{document}